\documentclass[twocolumn,prl,aps,superscriptaddress,showpacs]{revtex4-1}

\usepackage{mathptmx}
\usepackage{subfigure}
\usepackage{psfrag,graphicx}
\usepackage{dcolumn}
\usepackage{amsmath,amssymb}
\usepackage{bm}
\usepackage{color}
\usepackage{latexsym}
\usepackage{epstopdf}
\usepackage{color}
\usepackage[english]{babel}
\usepackage{latexsym}

\usepackage{subfigure}
\usepackage{amsfonts}
\usepackage{bm}
\usepackage{natbib}

\usepackage{appendix}

\definecolor{mygrey}{gray}{0.35}
\definecolor{myblue}{rgb}{0.2,0.2,0.8}
\definecolor{myzard}{cmyk}{0,0,0.05,0}
\definecolor{mywhite}{rgb}{1,1,1}
\definecolor{mywhite}{rgb}{1,1,1}
\definecolor{myred}{rgb}{1,0.,0.3}

\usepackage[colorlinks=true,citecolor=myblue,linkcolor=myred]{hyperref}

\usepackage{psfrag,graphicx} 
\usepackage{subfigure} 
\usepackage{amsmath} 
\usepackage{amssymb} 
\usepackage{amsfonts}
\usepackage{bm}
\usepackage{natbib}
\usepackage{epstopdf}
\usepackage{appendix}

\begin{document}


\title{Inducing Non-Classical Lasing Via Periodic Drivings in Circuit Quantum Electrodynamics}

 \author{Carlos Navarrete-Benlloch} 
 \affiliation{Max-Planck-Institut f\"ur Quantenoptik, Hans-Kopfermann-strasse 1, 85748 Garching, Germany}

 \author{Juan Jos\'e Garc\'{\i}a-Ripoll}
 \affiliation{Instituto de F\'{\i}sica Fundamental, IFF-CSIC, Serrano 113-bis, Madrid E-28006, Spain}

 \author{Diego Porras}
 \affiliation{Department of Physics and Astronomy, University of Sussex, Falmer, Brighton BN19QH, UK}
 \affiliation{Departamento de F\'isica Te\'orica I, Universidad Complutense, 28040 Madrid, Spain}

\date{\today}

\begin{abstract}
We show how a pair of superconducting qubits coupled to a microwave cavity mode can be used to engineer a
single-atom laser that emits light into a non-classical state. 
Our scheme relies on the dressing of the qubit-field
coupling by periodic modulations of the qubit energy. In the dressed basis, the radiative
decay of the first qubit becomes an effective incoherent pumping mechanism that injects energy into the system, hence turning dissipation to our advantage.
A second, auxiliary qubit is used to shape the decay within the cavity, in such a way that lasing occurs in a squeezed basis of the cavity mode. 
We characterize the system both by mean-field theory and exact calculations. 
Our work may find applications in the generation of squeezing and entanglement in circuit QED, as well as in the study of dissipative many-body phase transitions.
\end{abstract}

\maketitle

{\it Introduction.--}
Recent progress in experimental solid-state quantum optics has led to exciting possibilities for the control of quantum states of the electromagnetic field. Circuit quantum electrodynamics (QED) \cite{Blais04,Wallraff04nat} is one of such new platforms, and can be seen as the microwave counterpart of cavity QED, with optical cavities and atoms replaced, respectively, by linear and nonlinear superconducting circuits. The latter are usually referred to as `artificial atoms' or `superconducting qubits'. In circuit QED single emitters are placed permanently, and different quantum-optical elements can be combined by fabrication. The field emitted by those devices can be integrated into circuits in the form of itinerant fields, and hence, new ideas for generating quantum photonic states are of major importance for applications of this emerging field.

In recent years various experiments have shown that lasing by a single qubit is possible in this scenario \cite{Astafiev07nat,Grajcar08}, while at the same time the generation of squeezed states of the field via Josephson parametric amplifiers has taken a lot of attention \cite{Castellanos-Beltran08natphys,Mallet11,Murch13}. 
Here we propose a scheme that is motivated by two of 
the main advantages that circuit QED offers with respect to their optical counterparts:
(a) The transition frequency of superconducting qubits is in the microwave domain. Thus, one can modulate the system parameters with rates and amplitudes comparable to the transition energy. 
This opens up the way to a versatile control of qubit-field couplings with periodic drivings \cite{Porras12prl}. 
In atomic systems, on the contrary, controlling matter-light interactions typically involves Raman transitions which rely on the atomic internal structure \cite{Hammerer10rmp}.
(b) Several cavities and dissipative elements can be permanently coupled to a single artificial atom. Thus, they provide us with an ideal toolbox for engineering dissipative processes \cite{Murch12} that would be very challenging to implement in atomic QED.

Those advantages can be fully exploited to design a dissipative phase transition into a {\it lasing phase} in which light is emitted into a {\it squeezed state}, that is, a {\it non-classical state} in the sense of Glauber \cite{Glauber63,Dodonov02}. In particular, we show that:
(i) By introducing a periodic energy driving of the qubit, we are able to induce an effective counter-rotating type interaction between this and the field, what turns the qubit relaxation into an effective population inversion mechanism, hence turning dissipation into something useful. This leads to single atom lasing into a classical, coherent state.
(ii) A bi-periodic driving allows us to shape the qubit-field interaction such that photons are emitted into a squeezed photonic mode. 
A mean-field description of this problem allows us to predict a lasing transition. Surprisingly, if decay occurs by normal leakage of photons out of the cavity, dissipation still drives the system into a classical lasing phase.
(iii) An additional qubit can be used to induce a photon decay mechanism that cools the EM field into a squeezed vacuum \cite{Cirac93,Porras12prl}. The joint action of that cooling process and the emission of light into a squeezed mode, yields {\it lasing into a squeezed state}.
(iv) Our ideas can be implemented in circuit QED setups with state-of-the-art experimental parameters, thus leading
to a scheme that goes beyond single-atom lasing into coherent states in atomic \cite{McKeever03nat} or solid-state \cite{Xie07prl,Nomura10natphys} systems.

In this work numerical solutions of the master equation allow us to characterize the steady-state of the system and we show how finite-size effects modify the mean-field predictions. In addition to applications related to bright sources of squeezed or entangled light, our work paves the way to the study of dissipative phase transitions in mesoscopic lattice QED systems \cite{Schmidt13}, since many-qubit extensions of our scheme \cite{Quijandria13} pose an intriguing many-body problem where strongly correlated phenomena could be analyzed.

\begin{figure*}[t]
\includegraphics[width=0.9\textwidth]{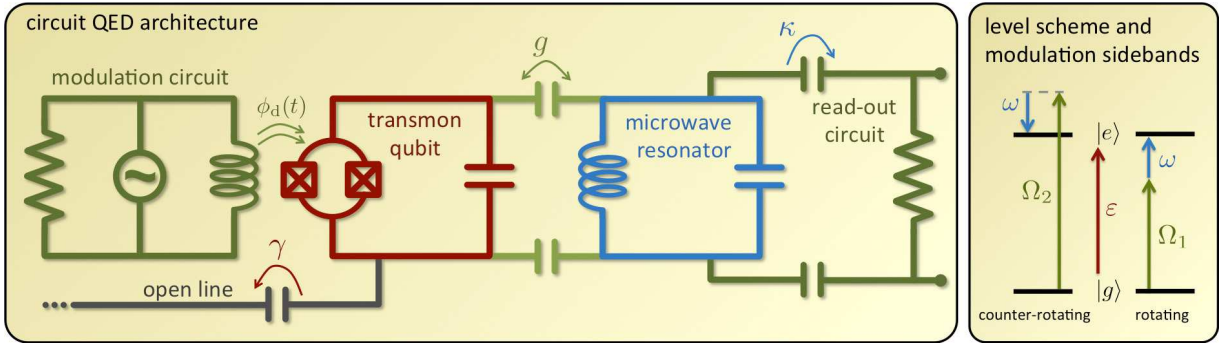}
\caption{(Left panel) Schematic proposal for a circuit QED architecture of the system: A superconducting qubit (transmon in the example) is capacitively coupled to an LC resonator of frequency $\omega_a$, while its transition frequency between the ground $|g\rangle$ and excited $|e\rangle$ states is modulated via the flux generated by an external circuit; the qubit is additionally coupled to an open transmission line which acts as an environment to which it can radiate excitations, while the resonator is coupled to a read-out circuit which acts as a dissipative channel for it. (Right panel) Scheme of the frequencies involved in the system: a bi-periodic modulation with frequencies matching the lower and upper sidebands of the qubit-resonator system allows for independently tune the relative amplitudes of the rotating and counter-rotating processes in which the qubit excitation is accompanied by the absorption or emission of a photon, respectively, see Eq. (\ref{Hint}).}
\label{fig1}
\end{figure*}

{\it Single artificial atom and cavity system.--}
As shown schematically in Fig. \ref{fig1}, we consider one mode of a cavity coupled to an artificial atom whose transition frequency is modulated in time. Such a system is described by a time-dependent Hamiltonian $H(t) = H_0 + H_{\rm int} + H_{\rm d}(t) $, with (we set $\hbar = 1 $)
\begin{eqnarray}
H_0 &=& \omega a^\dagger a + \frac{\epsilon}{2} \sigma_z ,
\ \
H_{\rm int} = g (a + a^\dagger) \sigma_x, \nonumber
\\
H_{\rm d}(t) &=& 
\sum_{j=1}^{n_{\rm d}} \Omega_j \eta_j \cos(\Omega_j t) \sigma_z,
\end{eqnarray}
where we have assumed that the modulation is multi-periodic. $\omega$ is the cavity frequency and $a$ the corresponding annihilation operator; $\sigma_{z,x}$ are the Pauli operators associated to the qubit, and $H_{\rm d}(t)$ describes $n_{\rm d}$ periodic drivings with frequencies $\Omega_j$ and normalized amplitudes $\eta_j$. 

Additionally, we consider two dissipative channels, one describing the radiative decay of the qubit to an open transmission line at rate $\gamma$, and another for the cavity losses at rate $\kappa$. Along the Letter we employ the notation  ${\cal L}_{\{ O, \Gamma \}}[\rho] = \Gamma \left( 2O \rho O^\dagger - O^\dagger O \rho-  \rho O^\dagger O \right)$, such that the master equation governing the evolution of the system's state $\rho$ reads
\begin{equation}
\dot{\rho} = -i [ H(t), \rho ] + \mathcal{L}_{\{\sigma,\gamma\}}[\rho]+\mathcal{L}_{\{a,\kappa\}}[\rho].
\label{GenMasterEq}
\end{equation}

We will be considering a far-off resonant and weak coupling regime ($g \ll \epsilon, \omega, |\epsilon-\omega|$), such that in the absence of driving the steady-state corresponds to the trivial photon vacuum. However, we show below that by switching on an appropriate modulation $H_{\rm d}(t)$, energy can be injected into the system, driving it into a lasing regime.

{\it Shaping the qubit-field interaction.--}
Consider a bi-chromatic driving ($n_\mathrm{d}=2$) modulating at the upper and lower sidebands of the qubit-cavity system, $\Omega_{1,2 }=\epsilon \mp \omega$, see Fig. \ref{fig1}. Moving to an interaction picture with respect to $H_0 + H_{\rm d}(t)$, we show in the supplemental material \cite{SupMat} that the system dynamics is well captured by the time-independent Hamiltonian
\begin{equation}
\tilde{H} = -\tilde{g} \left( u a^\dagger + v a \right) \sigma^\dagger + {\rm H.c.},
\label{Hint}
\end{equation}
where we have defined the parameters $u = J_0(2 \eta_1) J_1(2 \eta_2)/N$ and $v = J_0(2 \eta_2) J_1(2 \eta_1)/N$, which satisfy the Bogoliubov relation $|u^2 - v^2| = 1$ with the definition $N^2=|J_0^2(2 \eta_1) J_1^2(2 \eta_2) - J_0^2(2 \eta_2) J_1^2(2 \eta_1)|$ being $J_m(z)$ the Bessel function of order $m$, as well as a renormalized coupling $\tilde{g} = g N$. Note that we are describing here the renormalization of the qubit-field interaction by photon-assisted tunneling in a non-perturbative regime with respect to the modulation amplitudes \cite{Hanggi_review}. Hence, we see that a bi-periodic modulation allows us to tune the relative weights of the rotating and counter-rotating terms of the qubit-field interaction, what we exploit in the following to generate lasing to coherent or squeezed states.

{\it Single-qubit lasing.--}
Consider first the simple case $\eta_1=0$, in which we drive the qubit with a single frequency ($u=1,v=0$). In this case the qubit is coupled to the cavity mode through a counter-rotating type interaction, so that the master equation of the system reads
\begin{equation}
\dot{\rho} = ig [ a^\dagger \sigma^\dagger + a \sigma , \rho ] + {\cal L}_{\{ \sigma,\gamma \}}[\rho] + {\cal L}_{ \{a,\kappa \} }[\rho].
\label{MasterEqSQL}
\end{equation}
By using the transformation $\sigma\leftrightarrow\sigma^\dagger$, we see that the qubit relaxation is related to an effective spin-pumping mechanism together with a co-rotating atom-light coupling. This leads to our first result: the pumping provided via the periodic driving induces a single-qubit lasing mechanism. Eq. (\ref{MasterEqSQL}) has been studied in previous works, and mean-field theory predicts a lasing transition that depends on the cooperativity parameter $C=g^2/\gamma\kappa$. If $C\gg1$ and the inversion rate is much faster than the cavity losses, $\gamma\gg\kappa$, the steady state of the cavity consists in a coherent state with a random phase. 

{\it Engineering nonclassical lasing.--}
Let us now consider the situation in which both driving amplitudes $\eta_{1,2}$ are nonzero, so that the qubit is coupled to a squeezed mode $A = u a + v a^\dagger$ instead of the original cavity mode $a$. Choosing $|u|>|v|$, the interaction takes again a counter-rotating form $\tilde{H} = -\tilde{g} (A^\dagger \sigma^\dagger + A \sigma )$.
This seems to suggest lasing into the squeezed mode $A$, and thus emission of a bright squeezed state of light. 
However, we show below that a careful study of the master equation shows that this is not the case, since losses still take place by photon decay in the the original cavity mode basis, $a$, through the term
$\mathcal{L}_{a,\kappa}$ in Eq. (\ref{GenMasterEq}).
We prove in the following that in order to achieve lasing in the squeezed mode, $A$, cavity decay has to occur in that basis.
We thus introduce a second, auxiliary qubit, that will be used to control the photon decay in the cavity, following the ideas introduced in \cite{Porras12prl}. 
We assume that the auxiliary qubit is controlled by the same driving parameters, except for an exchange of the amplitudes $\eta_1\leftrightarrow\eta_2$ which makes $|u|<|v|$, such that  one effectively generates the rotating-type interaction 
$\tilde{H}' = -\tilde{g}' (A^\dagger {\sigma'}^\dagger + A \sigma' )$, where $\sigma'$ and $\tilde{g}'$ correspond to operators and couplings of the auxiliary qubit, respectively. 
The latter has a decay rate $\gamma'$, such that in the limit $\gamma'\gg\tilde{g}' \sqrt{\langle a^\dagger a \rangle}$, it can can be adiabatically eliminated. Finally, we obtain the master equation,
\begin{equation}
\dot{\rho} = i\tilde{g} [ A^\dagger \sigma^\dagger + A \sigma , \rho ] + {\cal L}_{\{ \sigma,\gamma \}}[\rho] + {\cal L}_{ \{ a,\kappa \}}[\rho]+\mathcal{L}_{\{A,\kappa \tilde{C}'\}}[\rho].
\label{MasterEq}
\end{equation}
If condition $\tilde{C}'=\tilde{g}'^2/\gamma'\kappa\gg v^2$ is met, the effective dissipator in the squeezed mode $\mathcal{L}_{\{A,\kappa \tilde{C}'\}}$ dominates the natural cavity dissipation ${\cal L}_{ \{ a,\kappa \}}$. Here we expect the system to behave as a single-qubit laser for the squeezed mode $A$, and hence as a nonclassical laser with respect to the original cavity mode $a$.

\begin{figure*}[t]
\includegraphics[width=\textwidth]{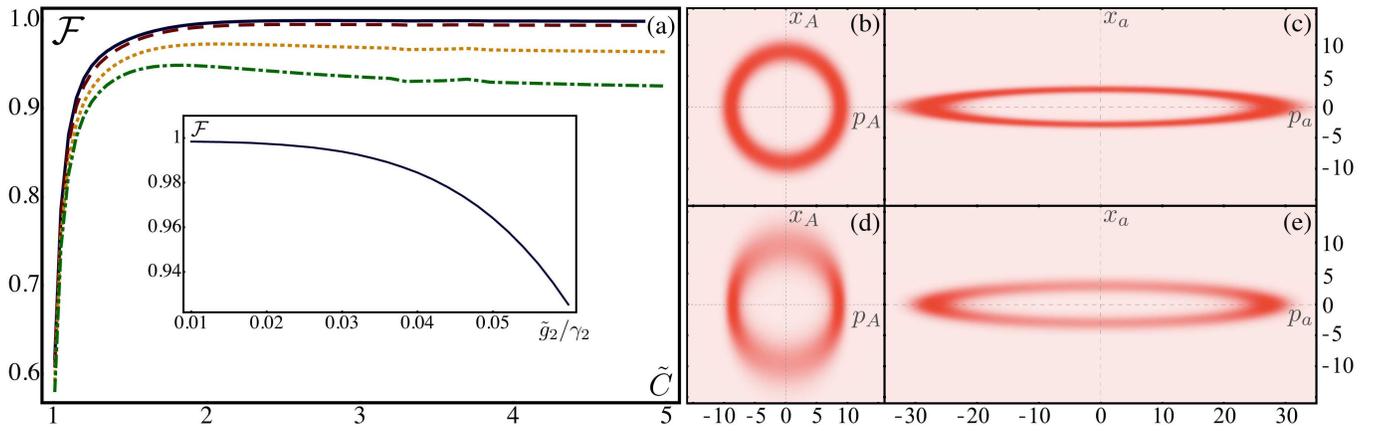}
\caption{(a) Fidelity between the mean-field ansatz (\ref{ss.A}) and the exact steady state of the system, as a function of the cooperativity $\tilde{C}$, for $\kappa/\gamma=0.02$, $r\approx 1.15$ (90\% of quadrature squeezing), and $\tilde{C}'=10$. The solid, blue curve corresponds to the exact steady state of Eq. (\ref{MasterEq}), while in the other curves the effect of the second qubit is considered for $\tilde{g}'/\gamma'=0.02$ (dashed red), $0.05$ (dotted yellow), and $0.07$ (dashed-dotted green); the inset shows the fidelity as a function of the ratio $\tilde{g}'/\gamma'$, fixing the parameters as in the main plot, plus $\tilde{C}=5$. We also show density plots of the Wigner functions corresponding to the steady-state of (\ref{MasterEq}) for $\tilde{C}=5$ and $\kappa/\gamma=0.02$, and two values of $\tilde{C}'$, 10 (b,c) and 0.01 (d,e), in which the states are well approximated by (\ref{AnsatzLimit1}) and  (\ref{AnsatzLimit2}) respectively.}
\label{fig2}
\end{figure*}

In order to get an approximate description of the steady-state predicted by this master equation, we apply a mean-field approximation in which $\rho$ is assumed to be separable in the qubit-field subspaces \cite{Breuer_book}. Defining the expectation values $F= \langle A \rangle$, $S = i \langle \sigma \rangle^*$, and $D = - \langle \sigma_z \rangle$, we get the nonlinear system of equations
\begin{eqnarray}
\dot{F} &=& - \kappa(1+\tilde{C}') F + \tilde{g} S, \ \ \dot{S} = -\gamma S + \tilde{g} D F, \nonumber \\
\dot{D} &=& - 2 \tilde{g} \left( S F^* + S^* F \right) - 2 \gamma \left( D - 1 \right),
\label{mf}
\end{eqnarray}
which are the so-called Maxwell-Bloch equations well known in laser physics \cite{Mandel95book,Breuer_book}. The steady-state solution of these equations predicts a lasing transition depending on the cooperativity parameter $\tilde{C} = \tilde{g}^2 / \gamma \kappa(1+\tilde{C}')$, which separates a trivial phase with 
$\bar{F}_A = \bar{S} = 0$ and $\bar{D} = 1$ (bar indicates steady-state values within the mean-field approximation) for $\tilde{C} < 1$, from a bright phase when $\tilde{C}>1$ in which 
\begin{equation}
\bar{F} = \sqrt{\frac{\gamma(\tilde{C}-1)}{2\kappa \tilde{C}}} e^{i\theta}, \ \ \bar{S} = \frac{\tilde{g}}{\tilde{C}\gamma} \bar{F}_A,  \ \  \bar{D} = \frac{1}{\tilde{C}},
\label{ClassSol}
\end{equation} 
where $\theta$ is an arbitrary phase not fixed by the equations. Note that deep into the lasing regime ($\tilde{C}\rightarrow \infty$) the number of (mean-field) photons depends solely on the ratio $\kappa/\gamma$.

The mean-field approximation allows us also to estimate the reduced steady state of the field $\rho_\mathrm{f} = {\rm tr}_{\rm qubit}\{\rho\}$. For this, we just use the fact that within this approximation the state is separable in the qubit-field subspaces, so that taking the partial trace of (\ref{MasterEq}), and using (\ref{ClassSol}), we get
\begin{equation}
(1+\tilde{C}')[\bar{F} A^\dagger - \bar{F}^* A, \bar{\rho}_\mathrm{f}]+{\cal L}_{ \{ u A - v A^\dagger, 1 \} }[\bar{\rho}_\mathrm{f}]+\mathcal{L}_{\{A,\tilde{C}'\}}[\bar{\rho}_\mathrm{f}] = 0,
\label{eff.LA}
\end{equation}
which, using the parametrization $\{u=\cosh r,v = \sinh r\}$ with $r \in [0,\infty[$ (we assume from now on $u,v>0$ without loss of generality), is easily shown to have the following Gaussian state as a solution (see the Supplemental Material \cite{SupMat,Weedbrook12}):
\begin{equation}
\rho_\mathrm{f}^\mathrm{G}(\bar{F},\tilde{C}',r)=D_{A}(\bar{F})S_{A}(\tilde{r})\rho_{\mathrm{th},A}(\tilde{n})S_{A}^\dagger(\tilde{r})D_{A}^\dagger(\bar{F}),
\label{GaussianMFsol}
\end{equation}
where we have defined the displacement and squeezing operators $D_A(\alpha) = \exp(\alpha A^\dagger-\alpha^*A)$ and $S_A(r) = \exp[r(A^{\dagger2}-A^2)/2]$, respectively, and the thermal state
\begin{equation}
\rho_{\mathrm{th},A}(N)=\sum_{n=0}^\infty\frac{N^n}{(1+N)^{1+n}}|n\rangle_{A}\langle n|,
\end{equation}
$|n\rangle_{A}$ referring to the Fock states associated to mode $A$, and where the squeezing parameter and thermal occupation number of the Gaussian state are given by 
\begin{eqnarray}
\tilde{r}&=&\ln\{[\exp(2r)+\tilde{C}']/[\exp(-2r)+\tilde{C}']\}/4,
\\
\tilde{n}&=&\left\{\sqrt{[\exp(2r)+\tilde{C}'][\exp(-2r)+\tilde{C}']/(1+\tilde{C}')^2}-1\right\}/2 \nonumber.
\end{eqnarray}
Then, taking into account that the mean-field solution (\ref{ClassSol}) assumes spontaneous symmetry breaking, whereas in reality the statistics over many realizations would show a random phase $\theta$, our mean-field ansatz is given by the mixture
\begin{equation}
\bar{\rho}_\mathrm{f}(|F|,\tilde{C}',r) = \int_0^{2 \pi} \frac{d\theta}{2 \pi} \rho_\mathrm{f}^\mathrm{G}(F,\tilde{C}',r).
\label{ss.A}
\end{equation}
This mean-field state is a generalization of the usual coherent-state mixture found in the laser \cite{Mandel95book,Scully97book,Walls94book}; below we discuss how well it describes the system compared to the exact steady state, but, before doing so, let us consider two physically relevant limits. First, the limit $\tilde{C}'\gg\exp(2r)$, in which $\tilde{r}=0$ and $\tilde{n}=0$, so that the ansatz can be written as
\begin{eqnarray}
\bar{\rho}_\mathrm{f}^{(1)} &=& \int_0^{2 \pi} \frac{d\theta}{2 \pi} D_{A}(\bar{F})|0\rangle_{A}\langle 0|D_{A}^\dagger(\bar{F})
\\
&=&\int_0^{2 \pi} \frac{d\theta}{2 \pi} S_{a}^\dagger(r)D_{a}(\bar{F})|0\rangle_{a}\langle 0|D_{a}^\dagger(\bar{F})S_{a}(r);\nonumber
\label{AnsatzLimit1}
\end{eqnarray}
we see that, as expected, in this limit the state is just a balanced mixture of all the coherent states of mode $A$ with the same mean-field amplitude $|\bar{F}|$, which is the ideal laser state. Hence, this is the limit in which our system works as a nonclassical laser, since this state corresponds to a mixture of squeezed coherent states in the basis of the original cavity mode (as shown explicitly after the second equality). The second limit we want to consider is $\tilde{C}'\rightarrow 0$, that is, the limit in which we do not add a second qubit to engineer dissipation in the squeezed mode $A$. In this case, $\tilde{n}=0$ again, but $\tilde{r}=r$, so that the mean-field ansatz can be written as
\begin{eqnarray}
\bar{\rho}_\mathrm{f}^{(2)} &=& \int_0^{2 \pi} \frac{d\theta}{2 \pi} D_{A}(\bar{F})S_{A}(r)|0\rangle_{A}\langle 0|S_{A}^\dagger(r)D_{A}^\dagger(\bar{F})
\\
&=&\int_0^{2 \pi} \frac{d\theta}{2 \pi} D_{a}(u\bar{F}-v\bar{F}^*)|0\rangle_{a}\langle 0|D_{a}^\dagger(u\bar{F}-v\bar{F}^*);\nonumber
\label{AnsatzLimit2}
\end{eqnarray}
this shows that without the help of the second qubit, the lasing process is still classical from the point of view of the original mode $a$, that is, the state is a mixture of coherent states.

Our laser works in a mesoscopic photon number regime in which the validity of the mean-field solution must be handled with care, and hence we proceed to study numerically the exact steady state of (\ref{MasterEq}). In the solid, blue curve of Fig. \ref{fig2}(a) we show the fidelity $\mathcal{F}$ between this exact steady state and the mean-field ansatz (\ref{ss.A}), as we move up into the lasing transition for 90\% of quadrature squeezing ($r\approx 1.15$), $\kappa/\gamma=0.02$, and $\tilde{C}'=10$ (similar curves are found for other values of $\tilde{C}'$). It can be appreciated how the mean-field ansatz adapts very well to the exact steady state above the lasing transition. In addition, in the rest of the curves of Fig. \ref{fig2}(a), we show the fidelity between the ansatz and the exact steady state of the system when the second qubit is not adiabatically eliminated, for different values of $\tilde{g}'/\gamma'$; we can appreciate that $\tilde{g}'/\gamma'\lesssim0.03$ is needed in order to achieve the lasing conditions we seek for. 

In order to characterize better the state in the different regimes, Figs. \ref{fig2}(b)-(e) show the Wigner functions corresponding to the limiting situations $\tilde{C}'\ll 1$ and $\tilde{C}'\gg v$, characterized by states (\ref{AnsatzLimit1}) and (\ref{AnsatzLimit2}), respectively---see \cite{SupMat,Cahill69,Brune92,Garraway92,Weedbrook12} for the details of their evaluation---. In particular, in Figs. \ref{fig2}(d) and \ref{fig2}(e) we plot these Wigner functions in the phase space of the original cavity mode $a$, which is what would be reconstructed in a tomography experiment along the lines of \cite{Mallet11,Hofheinz09nat}. Let us remark that, since all our Wigner functions are positive everywhere in the phase space formed by the quadratures $x_c=c^\dagger+c$ and $p_c=i(c^\dagger-c)$, where $c=A$ or $a$, these can be interpreted as just the joint probability distribution describing measurements of these observables.

{\it Physical implementation in circuit QED setups.--}
In order to make some connection with physical setups, let us propose a concrete circuit QED architecture with which it should be possible to test our ideas. This is sketched in Fig. \ref{fig1}: a transmon qubit \cite{Koch07,Schreier08} is capacitively coupled to an LC circuit, while its transition frequency is modulated via the flux generated by an external circuit which drives the Josephson loop. We take $\epsilon/2\pi=10\mathrm{GHz}$, $\omega/2\pi=4.5\mathrm{GHz}$, and $g/2\pi$ tunable up to $40\mathrm{MHz}$, which are common parameters in state of the art superconducting circuits \cite{Hofheinz09nat}. In addition, the qubit is strongly coupled to an open transmission line, what induces a relatively fast radiative decay rate $\gamma/2\pi=15\mathrm{MHz}$, while a read-out circuit induces a damping rate $\kappa/2\pi=30\mathrm{KHz}$ on the LC resonator. Single-qubit lasing with cooperativities and photon numbers up to $2000$ and $250$, respectively, can be achieved with this parameters. In order to generate the squeezed lasing proposed in the Letter, one could include a second qubit with $g'/2\pi$ tunable up to $70\mathrm{MHz}$ and a strong radiative decay $\gamma'/2\pi=250\mathrm{MHz}$, conditions in which its adiabatic elimination should be valid. As for the driving parameters, let us fix $\eta_2=0.2$, such that the corresponding physical modulation amplitude would be $\Omega_2 \eta_2 = 2.9\mathrm{GHz}$ which is quite reasonable. For this small normalized amplitude, we can approximate $\tanh r = v/u \approx \eta_1/\eta_2$, so that 90\% of quadrature squeezing ($r\approx1.15$) is obtained by choosing $\eta_1\approx 0.16$; then, and taking into account that the renormalized coupling can also be approximated by $\tilde{g}\approx g\sqrt{\eta_2^2-\eta_1^2}$ (similarly for $\tilde{g}'$), one can get up to cooperativities $\tilde{C}=5$ and $\tilde{C}'=10$, enough to see the phenomena introduced in the Letter.

{\it Conclusions and Outlook.--}
In this Letter we have shown how to engineer a single-atom laser that emits light into a non-classical state in a circuit QED scenario. Our scheme relies only on the modulation of the transition frequencies of two qubits with periodic drivings and exploits their radiative decay to our advantage: for one qubit it is turned into the effective population inversion mechanism needed for lasing, while for the other it allows engineering the cavity dissipation such that the lasing process becomes nonclassical. The generalization of our ideas to the generation of multi-mode squeezed and entangled states is straightforward, while the extension of our work to many qubits would allow studying strongly-correlated phenomena with circuit QED setups, providing the exciting possibility of preparing non-trivial many-body states dissipatively \cite{Diehl08,Verstraete09,Barreiro10,Barreiro11,Krauter11}. In addition to this, our laser works in a mesoscopic regime in which its response to a weak coherent signal could be studied to explore the physics of spontaneous symmetry breaking.

{\it Acknowledgments.-} We thank G\'eza Giedke for interesting suggestions concerning the characterization of the state of the system, Irfan Siddiqi, Steven J. Weber, and Kater W. Murch, for insight about the concrete circuit QED architecture and the corresponding system parameters, and Alfredo Luis for fruitful discussions.
C.N.-B. acknowledges the support from the Alexander von Humboldt Foundation through its Fellowship for Postdoctoral Researchers, and of the Future and Emerging Technologies (FET) programme within the Seventh Framework Programme for Research of the European Commission under the FET-Open grant agreement MALICIA, number FP7-ICT-265522.
J.J.G.-R acknowledges FET-Open project PROMISCE, CAM Research Consortium QUITEMAD (S2009-ESP-1594), and MINECO Project FIS2012-33022.
D.P. acknowledges RyC Contract No. Y200200074.

\begin{center}
\textbf{Supplemental material}
\end{center}

In this supplemental material we offer a detailed derivation of three points of the main Letter: (i) the time-independent Hamiltonian which captures the dynamics induced by the full time-dependent Hamiltonian modeling the
driven qubit-field system; (ii) the Gaussian-state solution of the field within the mean-field approximation; and (iii) the construction of the Wigner functions from the density matrices obtained numerically in the Fock basis.

\section{Effective time-independent Hamiltonian}

In the main Letter, we claimed that the dynamics induced by the time-dependent Hamiltonian $H(t)=H_{0}+H_{\mathrm{int}}+H_{\mathrm{d}}(t)$, with 
\begin{align}
H_0 & =\omega a^\dagger a+\frac{\varepsilon}{2}\sigma_z, \ \ H_\mathrm{int}=g(a+a^\dagger)\sigma_x,  \notag
\\
H_\mathrm{d}(t) & =\sum_{j=1}^{2}\Omega_j \eta_j \cos(\Omega_j t)\sigma_z,
\end{align}
is well captured by the time-independent one
\begin{equation}
\tilde{H}=-\tilde{g}\left(ua^\dagger+va\right)\sigma^\dagger+\mathrm{H.c.},  \label{Heffective}
\end{equation}
where $\tilde{g}$ is a renormalized coupling and the parameters $u$ and $v$ satisfy the Bogoliubov relation $|u^{2}-v^{2}|=1$, provided that one works far from the strong-coupling regime and off-resonance $(\omega,\varepsilon,\vert\varepsilon-\omega\vert \gg g)$, and chooses the upper and lower sideband modulations $\Omega_{1,2}=\varepsilon\mp\omega$. In this first section of the supplemental material we prove this statement rigorously.

To this aim, let us first move to the interaction picture defined by the transformation operator
\begin{eqnarray}
U_\mathrm{c}(t) &=& \exp\left[-iH_0 t - i\int_0^t d\tau H_\mathrm{d}(\tau)\right]
\\
&=& \exp \left[-i\omega t a^\dagger a - i\left(\frac{\varepsilon t}{2}+\sum_{j=1}^2 \eta_j\sin\Omega_j t\right) \sigma_z\right] ,  \notag
\end{eqnarray}
which transforms the state of the qubit-field system as $\rho \rightarrow\rho_\mathrm{I}=U_\mathrm{c}^\dagger\rho U_\mathrm{c}$, so that it evolves now according to the Hamiltonian
\begin{align}
H_\mathrm{I}&=U_\mathrm{c}^\dagger[H_0+H_\mathrm{d}(t)]U_\mathrm{c}- H_0 - H_\mathrm{d}(t)
\\
& =g\left\{ a\sigma \exp\left[-i\left(\omega t+\varepsilon t+\sum_{j=1}^{2} 2\eta_j \sin\Omega_j t\right)\right] \right.   \notag
\\
& \left. +a\sigma^\dagger\exp\left[-i\left(\omega t-\varepsilon t-\sum_{j=1}^2 2\eta_j\sin\Omega_j t\right)\right] \right\}+\mathrm{H.c.},  \notag
\end{align}
where we have used
\begin{eqnarray}
U_\mathrm{c}^\dagger a U_\mathrm{c} &=& a\exp(-i\omega t),
\\
U_\mathrm{c}^\dagger \sigma U_\mathrm{c} &=& \sigma\exp\left[-i\left(\varepsilon t+\sum_{j=1}^2 2\eta_j\sin\Omega_j t\right)\right].
\end{eqnarray}

The next step in the derivation consists in using the fact that the sine function is the generator of the Bessel functions, what means that%
\begin{equation}
\exp(2i\eta_j\sin\Omega_j t)=\sum_{n=-\infty}^{+\infty}J_{n}(2\eta_j)\exp(in\Omega_j t),  \label{SineGen}
\end{equation}
leading to the Hamiltonian
\begin{equation}
H_\mathrm{I}=\hbar g\left[\alpha(t)a\sigma^\dagger+\beta(t)a\sigma\right]+\mathrm{H.c.},  \label{Halphabeta}
\end{equation}
with
\begin{subequations}
\begin{align}
\alpha(t) & =\sum_{n_1,n_2=-\infty}^{+\infty}J_{n_1}(2\eta_1)J_{n_2}(2\eta_2)e^{-i\left(\omega-\varepsilon-n_1\Omega_1-n_2\Omega_2\right) t},
\\
\beta(t) & =\sum_{n_1,n_2=-\infty}^{+\infty}J_{n_1}(2\eta_1)J_{n_2}(2\eta_2)e^{-i\left(\omega+\varepsilon+n_1\Omega_1+n_2\Omega_2\right) t}.
\end{align}
\end{subequations}
This Hamiltonian has both rotating ($a\sigma^\dagger$) and counter-rotating ($a\sigma$) terms; however, these terms will contribute to the dynamics of the system only if some of the complex exponentials appearing in the definition of $\alpha(t)$ and $\beta(t)$ vary slowly compared to $g$ (rotating-wave approximation), that is, introducing $\Omega_{1,2}=\varepsilon \mp \omega $, the rotating term will
contribute only for $(m_1,m_2)$ such that 
\begin{equation}
\vert (1+m_1+m_2)\omega-(1+m_1-m_2)\varepsilon\vert \ll g, \label{Cond1}
\end{equation}
while the counter-rotating term will enter the dynamics only if
\begin{equation}
\vert (1-q_1+q_2)\omega+(1+q_1+q_2)\varepsilon\vert \ll g,
\end{equation}
for some combination $(q_1,q_2)$. It is possible to find exponentials which oscillate slow compared to $g$ both in $\alpha (t)$ and $\beta (t)$. In particular, provided the no multi-photon resonances are allowed within the coupling strength, that is
\begin{equation}
\vert m\varepsilon -n\omega \vert \gg g\text{ \ \ \ }\forall mn=1,2,...,  \label{Multi-Ph}
\end{equation}
only one term of $\alpha(t)$ and another of $\beta(t)$ survive, the ones with $(m_1=-1,m_2=0)$ and $(q_1=0,q_2=-1)$, respectively. Note
however that it is enough that condition (\ref{Multi-Ph}) holds for small $m$ and $n$, as if the multi-photon resonance occurs only for large ones, only high order Bessel functions kick in, and then the terms previously found are still the only ones which contribute to $\alpha(t)$ and $\beta(t)$ approximately. For example, for the frequencies chosen in the Letter, $\epsilon/2\pi=10\mathrm{GHz}$ and $\omega/2\pi=4.5\mathrm{GHz}$, the first multi-photon resonance that satisfies (\ref{Cond1}) is $(m_1=28,m_2=11)$, which gives a negligible contribution to $\alpha(t)$ unless the modulation amplitudes $\eta_j$ are extremely large.

Under such conditions, the Hamiltonian (\ref{Halphabeta}) takes the form
\begin{equation}
H_\mathrm{I}\approx g\left[J_{-1}(2\eta_1)J_0(2\eta_2)a+J_0(2\eta_1)J_{-1}(2\eta_2)a^\dagger\right] \sigma^\dagger+\mathrm{H.c.};
\end{equation}
now, using the property $J_{-1}(x)=-J_1(x)$, and defining the parameters
\begin{subequations}
\begin{eqnarray}
v &=&\frac{J_1(2\eta_1)J_0(2\eta_2)}{\sqrt{\left\vert J_1^2(2\eta_1)J_0^2(2\eta_2)-J_0^2(2\eta_1)J_1^2(2\eta_2)\right\vert}},
\\
u &=&\frac{J_0(2\eta_1)J_1(2\eta_2)}{\sqrt{\left\vert J_1^2(2\eta_1)J_0^2(2\eta_2)-J_0^2(2\eta_1)J_1^2(2\eta_2)\right\vert}},
\\
\bar{g} &=& g\sqrt{\left\vert J_1^2(2\eta_1)J_0^2(2\eta_2)-J_0^2(2\eta_1)J_1^2(2\eta_2)\right\vert},
\end{eqnarray}
\end{subequations}
we obtain the Hamiltonian (\ref{Heffective}) as we wanted to prove.

\section{Gaussian-state solution to the mean-field equation}

In this section we find the steady-state solution of the master equation
\begin{equation}
\dot{\rho}_\mathrm{f}=(1+\tilde{C}^\prime)[\bar{F}A^\dagger-\bar{F}^\ast A,\rho_\mathrm{f}]+\mathcal{L}_{\{a,1\}}[\rho_\mathrm{f}]+\mathcal{L}_{\{A,\tilde{C}^\prime\}}[\rho_\mathrm{f}], \label{MFmaster}
\end{equation}
with $a=A\cosh r-A^\dagger\sinh r$, which corresponds to the state of the cavity field within the mean-field approximation. Note that given any field operator $O$, we can find the evolution equation of its expectation value as
\begin{eqnarray}
\langle \dot{O}\rangle  &=&\mathrm{tr}\{O\dot{\rho}_\mathrm{f}\}=(1+C^\prime)\langle [O,\bar{F}A^\dagger-\bar{F}^\ast A]\rangle +\langle a^\dagger[O,a]\rangle   \notag
\\
&&+\langle [a^\dagger,O]a\rangle +\tilde{C}^\prime\langle A^\dagger[O,A]\rangle +\tilde{C}^\prime\langle [A^\dagger,O]A\rangle.
\label{ObsEvo}
\end{eqnarray}

Now, since equation (\ref{MFmaster}) is quadratic in annihilation and creation operators $(A,A^\dagger)$, its steady state $\bar{\rho}_\mathrm{f}$ is Gaussian (from now on the bar denotes steady-state values), meaning that it is completely characterized by its first and second moments \cite{Weedbrook12}. In particular, using (\ref{ObsEvo}) it is simple to find $\overline{\langle A\rangle}=\bar{F}$,
\begin{subequations}
\begin{eqnarray}
\overline{\langle A^\dagger A\rangle} &=& \vert\bar{F}\vert^2+\frac{\sinh^2 r}{1+\tilde{C}^\prime},
\\
\overline{\langle A^2\rangle} &=& \bar{F}^2+\frac{\sinh 2r}{2(1+\tilde{C}^\prime)}.
\end{eqnarray}
\end{subequations}
Defining the quadratures $x_A=A^\dagger+A$ and $p_A=i(A^\dagger-A)$, the vector operator $\mathbf{r}_A=\mathrm{col}(x_A,p_A)$, and the corresponding mean vector $\mathbf{d}_A=\langle\mathbf{r}_A\rangle$ and covariance matrix $V_A$ with elements $V_{A,jk}=\langle r_{A,j}r_{A,k}\rangle-\langle r_{A,j}\rangle \langle r_{A,k}\rangle$, we then get a state with Gaussian Wigner function
\begin{equation}
\bar{W}_\mathrm{f}(\mathbf{R}_A)=\frac{1}{2\pi\sqrt{\det\bar{V}_A}}e^{-(\mathbf{R}_A-\mathbf{\bar{d}}_A)^T\bar{V}_A^{-1}(\mathbf{R}_A-\mathbf{\bar{d}}_A)/2},  \label{GaussianWigner}
\end{equation}
where $\mathbf{R}_A=\mathrm{col}(X_A,P_A)$ are phase space variables associated to the quadrature operators (in the main Letter we kept the names $x_A$ and $p_A$ for these $c$-numbers in Fig. 2 for simplicity), and
\begin{subequations}
\begin{eqnarray}
\mathbf{\bar{d}}_A &=&2\mathrm{col}(\mathrm{Re}\{\bar{F}\},\mathrm{Im}\{\bar{F}\}),  \label{dA}
\\
\bar{V}_A &=&\frac{1}{\tilde{C}^\prime +1}
\begin{pmatrix}
\tilde{C}^\prime+e^{2r} & 0
\\ 
0 & \tilde{C}^\prime+e^{-2r}
\end{pmatrix}.  \label{VA}
\end{eqnarray}
\end{subequations}

In order to gain more insight, we are going to write this Gaussian state in a different manner. Concretely, it is well known that any single-mode Gaussian state can always be written in the form \cite{Weedbrook12}
\begin{equation}
\rho=D_A(\alpha)R_A(\varphi)S_A(\tilde{r})\rho_{\mathrm{th},A}(\tilde{n})S_A^\dagger(\tilde{r})R_A^\dagger(\varphi) D_A^\dagger(\alpha),  \label{GeneralGaussian}
\end{equation}
where we have defined the displacement $D_A(\alpha)=\exp(\alpha A^\dagger-\alpha^\ast A)$, phase-shift $R_A(\varphi)=\exp(i\varphi A^\dagger A)$, and squeezing $S_A(\tilde{r})=\exp[\tilde{r}(A^{\dagger 2}-A^2)/2]$ operators, as well as the thermal state $\rho_{\mathrm{th},A}(\tilde{n})$, which is a Gaussian state with zero mean vector and covariance matrix $V_{\mathrm{th},A}(\tilde{n})=(2\tilde{n}+1)I_{2\times 2}$. The displacement parameter $\alpha$ coincides with the mean of the state, what in our case means $\alpha=\bar{F}$, while no phase-shift is needed ($\varphi=0$) for a state with a diagonal covariance matrix as (\ref{VA}). On the other hand, since the entropy is invariant under unitary
transformations, and for a single-mode state it depends solely on the determinant of the covariance matrix \cite{Weedbrook12}, the thermal photon number parameter $\tilde{n}$ is found by matching the determinants of $V_{\mathrm{th},A}(\tilde{n})$ and $\bar{V}_A$, that is, $(2\tilde{n}+1)^2 = \det\bar{V}_A$. Finally, $S_A(\tilde{r})$ squeezes (anti-squeezes) the momentum (position) variance by a factor $e^{-2\tilde{r}}$ ($e^{2\tilde{r}}$), and hence the squeezing parameter $\tilde{r}$ is found from the asymmetry of the covariance matrix, that is, $\exp(4\tilde{r})=\bar{V}_{A,11}/\bar{V}_{A,22}$. Combining all these results, (\ref{GeneralGaussian}) is turned into the Gaussian state $\rho_\mathrm{f}^\mathrm{G}$ that we introduced in the Letter.

Note finally that, given the relation $A=S_a^\dagger(r)a S_a(r)$, the relation between the Fock basis of the squeezed and original cavity modes is $|n\rangle_A=S_a^\dagger(r)|n\rangle_a$, and hence the Gaussian state (\ref{GeneralGaussian}) with $\varphi=0$ can be written as
\begin{equation}
\rho=S_a^\dagger(r)D_a(\alpha)S_a(\tilde{r})\rho_{\mathrm{th},a}(\tilde{n})S_a^\dagger(\tilde{r})D_a^\dagger(\alpha)S_a(r),
\end{equation}
in the basis of the original cavity mode.

\section{Wigner functions from the density matrix}

All our numerics have been performed by using the Fock states $\{|n\rangle_A\}_{n=0,1,...,N_A}$ of the squeezed mode $A$ as the basis of the field's Hilbert space (truncated to a large enough photon number $N_A$), what gives us the reduced state of the cavity mode represented as $\rho_\mathrm{f}=\sum_{mn=0}^{N_A}\rho_{mn}^A|m\rangle_A\langle n|$. In
this section we explain how to find the Wigner functions in the phase space of both the squeezed mode $A$ and the original cavity mode $a$, starting from this representation of the state.

Let us write the polar form of the coordinate vector in the phase space of mode $A$ as $\mathbf{R}_A=R_A(\cos\phi_A,\sin\phi_A)$. Hence, based on the following result \cite{Cahill69,Brune92,Garraway92} for the Wigner function of the operator $|m\rangle_A\langle n|$:
\begin{eqnarray}
W_{mn}(R_A,\phi_A)=\frac{(-1)^n}{\pi} &&\sqrt{\frac{n!}{m!}}e^{i\phi_A(m-n)}R_A^{m-n}  \label{Wnm}
\\
&&\times L_n^{m-n}(R_A^2)e^{-R_A^2/2},  \notag
\end{eqnarray}
where $L_n^p(x)$ are the modified Laguerre polynomials and we have assumed $m\geq n$ (note that $W_{nm}=W_{mn}^\ast$), we get
\begin{equation}
W_A(R_A,\phi_A)=\sum_{mn=0}^{N_A}\rho_{mn}^A W_{mn}(R_A,\phi_A),  \label{Wsinglemode}
\end{equation}
which gives us the desired relation between the density matrix $\rho^A$ and the Wigner function $W_A(\mathbf{R}_A)$ in the phase space of the squeezed mode $A$.

On the other hand, in order to find the Wigner function in the phase space of the original cavity mode $a$, we just use the fact that $A=S_a^\dagger(r)a S_a(r)$ is equivalent to the symplectic transformation \cite{Weedbrook12} $\mathbf{R}_A=\mathcal{S}(r)\mathbf{R}_a$ between the corresponding phase spaces, with $\mathcal{S}(r)=\mathrm{diag}(e^r,e^{-r})$. Hence, given the Wigner function evaluated with (\ref{Wsinglemode}) in the phase space of mode $A$, the Wigner function in the phase space of mode $a$ is found as $W_a(\mathbf{R}_a)=W_A[\mathcal{S}(r)\mathbf{R}_a]$.

\bibliographystyle{apsrev4-1}
\bibliography{references_ncl}

\begin{thebibliography}{36}%
\makeatletter
\providecommand \@ifxundefined [1]{%
 \@ifx{#1\undefined}
}%
\providecommand \@ifnum [1]{%
 \ifnum #1\expandafter \@firstoftwo
 \else \expandafter \@secondoftwo
 \fi
}%
\providecommand \@ifx [1]{%
 \ifx #1\expandafter \@firstoftwo
 \else \expandafter \@secondoftwo
 \fi
}%
\providecommand \natexlab [1]{#1}%
\providecommand \enquote  [1]{``#1''}%
\providecommand \bibnamefont  [1]{#1}%
\providecommand \bibfnamefont [1]{#1}%
\providecommand \citenamefont [1]{#1}%
\providecommand \href@noop [0]{\@secondoftwo}%
\providecommand \href [0]{\begingroup \@sanitize@url \@href}%
\providecommand \@href[1]{\@@startlink{#1}\@@href}%
\providecommand \@@href[1]{\endgroup#1\@@endlink}%
\providecommand \@sanitize@url [0]{\catcode `\\12\catcode `\$12\catcode
  `\&12\catcode `\#12\catcode `\^12\catcode `\_12\catcode `\%12\relax}%
\providecommand \@@startlink[1]{}%
\providecommand \@@endlink[0]{}%
\providecommand \url  [0]{\begingroup\@sanitize@url \@url }%
\providecommand \@url [1]{\endgroup\@href {#1}{\urlprefix }}%
\providecommand \urlprefix  [0]{URL }%
\providecommand \Eprint [0]{\href }%
\providecommand \doibase [0]{http://dx.doi.org/}%
\providecommand \selectlanguage [0]{\@gobble}%
\providecommand \bibinfo  [0]{\@secondoftwo}%
\providecommand \bibfield  [0]{\@secondoftwo}%
\providecommand \translation [1]{[#1]}%
\providecommand \BibitemOpen [0]{}%
\providecommand \bibitemStop [0]{}%
\providecommand \bibitemNoStop [0]{.\EOS\space}%
\providecommand \EOS [0]{\spacefactor3000\relax}%
\providecommand \BibitemShut  [1]{\csname bibitem#1\endcsname}%
\let\auto@bib@innerbib\@empty
\bibitem [{\citenamefont {Blais}\ \emph {et~al.}(2004)\citenamefont {Blais},
  \citenamefont {Huang}, \citenamefont {Wallraff}, \citenamefont {Girvin},\
  and\ \citenamefont {Schoelkopf}}]{Blais04}%
  \BibitemOpen
  \bibfield  {author} {\bibinfo {author} {\bibfnamefont {A.}~\bibnamefont
  {Blais}}, \bibinfo {author} {\bibfnamefont {R.-S.}\ \bibnamefont {Huang}},
  \bibinfo {author} {\bibfnamefont {A.}~\bibnamefont {Wallraff}}, \bibinfo
  {author} {\bibfnamefont {S.~M.}\ \bibnamefont {Girvin}}, \ and\ \bibinfo
  {author} {\bibfnamefont {R.~J.}\ \bibnamefont {Schoelkopf}},\ }\href
  {\doibase 10.1103/PhysRevA.69.062320} {\bibfield  {journal} {\bibinfo
  {journal} {Phys. Rev. A}\ }\textbf {\bibinfo {volume} {69}},\ \bibinfo
  {pages} {062320} (\bibinfo {year} {2004})}\BibitemShut {NoStop}%
\bibitem [{\citenamefont {{Wallraff}}\ \emph {et~al.}(2004)\citenamefont
  {{Wallraff}}, \citenamefont {{Schuster}}, \citenamefont {{Blais}},
  \citenamefont {{Frunzio}}, \citenamefont {{Huang}}, \citenamefont {{Majer}},
  \citenamefont {{Kumar}}, \citenamefont {{Girvin}},\ and\ \citenamefont
  {{Schoelkopf}}}]{Wallraff04nat}%
  \BibitemOpen
  \bibfield  {author} {\bibinfo {author} {\bibfnamefont {A.}~\bibnamefont
  {{Wallraff}}}, \bibinfo {author} {\bibfnamefont {D.~I.}\ \bibnamefont
  {{Schuster}}}, \bibinfo {author} {\bibfnamefont {A.}~\bibnamefont {{Blais}}},
  \bibinfo {author} {\bibfnamefont {L.}~\bibnamefont {{Frunzio}}}, \bibinfo
  {author} {\bibfnamefont {R.-S.}\ \bibnamefont {{Huang}}}, \bibinfo {author}
  {\bibfnamefont {J.}~\bibnamefont {{Majer}}}, \bibinfo {author} {\bibfnamefont
  {S.}~\bibnamefont {{Kumar}}}, \bibinfo {author} {\bibfnamefont {S.~M.}\
  \bibnamefont {{Girvin}}}, \ and\ \bibinfo {author} {\bibfnamefont {R.~J.}\
  \bibnamefont {{Schoelkopf}}},\ }\href {\doibase 10.1038/nature02851}
  {\bibfield  {journal} {\bibinfo  {journal} {Nature}\ }\textbf {\bibinfo
  {volume} {431}},\ \bibinfo {pages} {162} (\bibinfo {year}
  {2004})}\BibitemShut {NoStop}%
\bibitem [{\citenamefont {{Astafiev}}\ \emph {et~al.}(2007)\citenamefont
  {{Astafiev}}, \citenamefont {{Inomata}}, \citenamefont {{Niskanen}},
  \citenamefont {{Yamamoto}}, \citenamefont {{Pashkin}}, \citenamefont
  {{Nakamura}},\ and\ \citenamefont {{Tsai}}}]{Astafiev07nat}%
  \BibitemOpen
  \bibfield  {author} {\bibinfo {author} {\bibfnamefont {O.}~\bibnamefont
  {{Astafiev}}}, \bibinfo {author} {\bibfnamefont {K.}~\bibnamefont
  {{Inomata}}}, \bibinfo {author} {\bibfnamefont {A.~O.}\ \bibnamefont
  {{Niskanen}}}, \bibinfo {author} {\bibfnamefont {T.}~\bibnamefont
  {{Yamamoto}}}, \bibinfo {author} {\bibfnamefont {Y.~A.}\ \bibnamefont
  {{Pashkin}}}, \bibinfo {author} {\bibfnamefont {Y.}~\bibnamefont
  {{Nakamura}}}, \ and\ \bibinfo {author} {\bibfnamefont {J.~S.}\ \bibnamefont
  {{Tsai}}},\ }\href {\doibase 10.1038/nature06141} {\bibfield  {journal}
  {\bibinfo  {journal} {Nature}\ }\textbf {\bibinfo {volume} {449}},\ \bibinfo
  {pages} {588} (\bibinfo {year} {2007})}\BibitemShut {NoStop}%
\bibitem [{\citenamefont {Grajcar}\ \emph {et~al.}(2008)\citenamefont
  {Grajcar}, \citenamefont {van~der Ploeg}, \citenamefont {Izmalkov},
  \citenamefont {Il/'ichev}, \citenamefont {Meyer}, \citenamefont {Fedorov},
  \citenamefont {Shnirman},\ and\ \citenamefont {Sch\"on}}]{Grajcar08}%
  \BibitemOpen
  \bibfield  {author} {\bibinfo {author} {\bibfnamefont {M.}~\bibnamefont
  {Grajcar}}, \bibinfo {author} {\bibfnamefont {S.~H.~W.}\ \bibnamefont
  {van~der Ploeg}}, \bibinfo {author} {\bibfnamefont {A.}~\bibnamefont
  {Izmalkov}}, \bibinfo {author} {\bibfnamefont {E.}~\bibnamefont {Il/'ichev}},
  \bibinfo {author} {\bibfnamefont {H.-G.}\ \bibnamefont {Meyer}}, \bibinfo
  {author} {\bibfnamefont {A.}~\bibnamefont {Fedorov}}, \bibinfo {author}
  {\bibfnamefont {A.}~\bibnamefont {Shnirman}}, \ and\ \bibinfo {author}
  {\bibfnamefont {G.}~\bibnamefont {Sch\"on}},\ }\href {\doibase
  10.1038/nphys1019} {\bibfield  {journal} {\bibinfo  {journal} {Nat. Phys.}\
  }\textbf {\bibinfo {volume} {4}},\ \bibinfo {pages} {612} (\bibinfo {year}
  {2008})}\BibitemShut {NoStop}%
\bibitem [{\citenamefont {{Castellanos-Beltran}}\ \emph
  {et~al.}(2008)\citenamefont {{Castellanos-Beltran}}, \citenamefont {{Irwin}},
  \citenamefont {{Hilton}}, \citenamefont {{Vale}},\ and\ \citenamefont
  {{Lehnert}}}]{Castellanos-Beltran08natphys}%
  \BibitemOpen
  \bibfield  {author} {\bibinfo {author} {\bibfnamefont {M.~A.}\ \bibnamefont
  {{Castellanos-Beltran}}}, \bibinfo {author} {\bibfnamefont {K.~D.}\
  \bibnamefont {{Irwin}}}, \bibinfo {author} {\bibfnamefont {G.~C.}\
  \bibnamefont {{Hilton}}}, \bibinfo {author} {\bibfnamefont {L.~R.}\
  \bibnamefont {{Vale}}}, \ and\ \bibinfo {author} {\bibfnamefont {K.~W.}\
  \bibnamefont {{Lehnert}}},\ }\href {\doibase 10.1038/nphys1090} {\bibfield
  {journal} {\bibinfo  {journal} {Nat. Phys.}\ }\textbf {\bibinfo {volume}
  {4}},\ \bibinfo {pages} {929} (\bibinfo {year} {2008})}\BibitemShut {NoStop}%
\bibitem [{\citenamefont {Mallet}\ \emph {et~al.}(2011)\citenamefont {Mallet},
  \citenamefont {Castellanos-Beltran}, \citenamefont {Ku}, \citenamefont
  {Glancy}, \citenamefont {Knill}, \citenamefont {Irwin}, \citenamefont
  {Hilton}, \citenamefont {Vale},\ and\ \citenamefont {Lehnert}}]{Mallet11}%
  \BibitemOpen
  \bibfield  {author} {\bibinfo {author} {\bibfnamefont {F.}~\bibnamefont
  {Mallet}}, \bibinfo {author} {\bibfnamefont {M.~A.}\ \bibnamefont
  {Castellanos-Beltran}}, \bibinfo {author} {\bibfnamefont {H.~S.}\
  \bibnamefont {Ku}}, \bibinfo {author} {\bibfnamefont {S.}~\bibnamefont
  {Glancy}}, \bibinfo {author} {\bibfnamefont {E.}~\bibnamefont {Knill}},
  \bibinfo {author} {\bibfnamefont {K.~D.}\ \bibnamefont {Irwin}}, \bibinfo
  {author} {\bibfnamefont {G.~C.}\ \bibnamefont {Hilton}}, \bibinfo {author}
  {\bibfnamefont {L.~R.}\ \bibnamefont {Vale}}, \ and\ \bibinfo {author}
  {\bibfnamefont {K.~W.}\ \bibnamefont {Lehnert}},\ }\href {\doibase
  10.1103/PhysRevLett.106.220502} {\bibfield  {journal} {\bibinfo  {journal}
  {Phys. Rev. Lett.}\ }\textbf {\bibinfo {volume} {106}},\ \bibinfo {pages}
  {220502} (\bibinfo {year} {2011})}\BibitemShut {NoStop}%
\bibitem [{\citenamefont {Murch}\ \emph {et~al.}(2013)\citenamefont {Murch},
  \citenamefont {Weber}, \citenamefont {Beck}, \citenamefont {Ginossar},\ and\
  \citenamefont {Siddiqi}}]{Murch13}%
  \BibitemOpen
  \bibfield  {author} {\bibinfo {author} {\bibfnamefont {K.~W.}\ \bibnamefont
  {Murch}}, \bibinfo {author} {\bibfnamefont {S.~J.}\ \bibnamefont {Weber}},
  \bibinfo {author} {\bibfnamefont {K.~M.}\ \bibnamefont {Beck}}, \bibinfo
  {author} {\bibfnamefont {E.}~\bibnamefont {Ginossar}}, \ and\ \bibinfo
  {author} {\bibfnamefont {I.}~\bibnamefont {Siddiqi}},\ }\href {\doibase
  10.1038/nature12264} {\bibfield  {journal} {\bibinfo  {journal} {Nature}\
  }\textbf {\bibinfo {volume} {499}},\ \bibinfo {pages} {62} (\bibinfo {year}
  {2013})}\BibitemShut {NoStop}%
\bibitem [{\citenamefont {Porras}\ and\ \citenamefont
  {Garc\'ia-Ripoll}(2012)}]{Porras12prl}%
  \BibitemOpen
  \bibfield  {author} {\bibinfo {author} {\bibfnamefont {D.}~\bibnamefont
  {Porras}}\ and\ \bibinfo {author} {\bibfnamefont {J.~J.}\ \bibnamefont
  {Garc\'ia-Ripoll}},\ }\href {\doibase 10.1103/PhysRevLett.108.043602}
  {\bibfield  {journal} {\bibinfo  {journal} {Phys. Rev. Lett.}\ }\textbf
  {\bibinfo {volume} {108}},\ \bibinfo {eid} {043602} (\bibinfo {year}
  {2012})}\BibitemShut {NoStop}%
\bibitem [{\citenamefont {Hammerer}\ \emph {et~al.}(2010)\citenamefont
  {Hammerer}, \citenamefont {S\o{}rensen},\ and\ \citenamefont
  {Polzik}}]{Hammerer10rmp}%
  \BibitemOpen
  \bibfield  {author} {\bibinfo {author} {\bibfnamefont {K.}~\bibnamefont
  {Hammerer}}, \bibinfo {author} {\bibfnamefont {A.~S.}\ \bibnamefont
  {S\o{}rensen}}, \ and\ \bibinfo {author} {\bibfnamefont {E.~S.}\ \bibnamefont
  {Polzik}},\ }\href {\doibase 10.1103/RevModPhys.82.1041} {\bibfield
  {journal} {\bibinfo  {journal} {Rev. Mod. Phys.}\ }\textbf {\bibinfo {volume}
  {82}},\ \bibinfo {pages} {1041} (\bibinfo {year} {2010})}\BibitemShut
  {NoStop}%
\bibitem [{\citenamefont {Murch}\ \emph {et~al.}(2012)\citenamefont {Murch},
  \citenamefont {Vool}, \citenamefont {Zhou}, \citenamefont {Weber},
  \citenamefont {Girvin},\ and\ \citenamefont {Siddiqi}}]{Murch12}%
  \BibitemOpen
  \bibfield  {author} {\bibinfo {author} {\bibfnamefont {K.~W.}\ \bibnamefont
  {Murch}}, \bibinfo {author} {\bibfnamefont {U.}~\bibnamefont {Vool}},
  \bibinfo {author} {\bibfnamefont {D.}~\bibnamefont {Zhou}}, \bibinfo {author}
  {\bibfnamefont {S.~J.}\ \bibnamefont {Weber}}, \bibinfo {author}
  {\bibfnamefont {S.~M.}\ \bibnamefont {Girvin}}, \ and\ \bibinfo {author}
  {\bibfnamefont {I.}~\bibnamefont {Siddiqi}},\ }\href {\doibase
  10.1103/PhysRevLett.109.183602} {\bibfield  {journal} {\bibinfo  {journal}
  {Phys. Rev. Lett.}\ }\textbf {\bibinfo {volume} {109}},\ \bibinfo {pages}
  {183602} (\bibinfo {year} {2012})}\BibitemShut {NoStop}%
\bibitem [{\citenamefont {Glauber}(1963)}]{Glauber63}%
  \BibitemOpen
  \bibfield  {author} {\bibinfo {author} {\bibfnamefont {R.~J.}\ \bibnamefont
  {Glauber}},\ }\href {\doibase 10.1103/PhysRev.131.2766} {\bibfield  {journal}
  {\bibinfo  {journal} {Phys. Rev.}\ }\textbf {\bibinfo {volume} {131}},\
  \bibinfo {pages} {2766} (\bibinfo {year} {1963})}\BibitemShut {NoStop}%
\bibitem [{\citenamefont {Dodonov}(2002)}]{Dodonov02}%
  \BibitemOpen
  \bibfield  {author} {\bibinfo {author} {\bibfnamefont {V.~V.}\ \bibnamefont
  {Dodonov}},\ }\href {http://stacks.iop.org/1464-4266/4/i=1/a=201} {\bibfield
  {journal} {\bibinfo  {journal} {J. Opt. B: Quantum Semicl. Opt.}\ }\textbf
  {\bibinfo {volume} {4}},\ \bibinfo {pages} {R1} (\bibinfo {year}
  {2002})}\BibitemShut {NoStop}%
\bibitem [{\citenamefont {Cirac}\ \emph {et~al.}(1993)\citenamefont {Cirac},
  \citenamefont {Parkins}, \citenamefont {Blatt},\ and\ \citenamefont
  {Zoller}}]{Cirac93}%
  \BibitemOpen
  \bibfield  {author} {\bibinfo {author} {\bibfnamefont {J.~I.}\ \bibnamefont
  {Cirac}}, \bibinfo {author} {\bibfnamefont {A.~S.}\ \bibnamefont {Parkins}},
  \bibinfo {author} {\bibfnamefont {R.}~\bibnamefont {Blatt}}, \ and\ \bibinfo
  {author} {\bibfnamefont {P.}~\bibnamefont {Zoller}},\ }\href {\doibase
  10.1103/PhysRevLett.70.556} {\bibfield  {journal} {\bibinfo  {journal} {Phys.
  Rev. Lett.}\ }\textbf {\bibinfo {volume} {70}},\ \bibinfo {pages} {556}
  (\bibinfo {year} {1993})}\BibitemShut {NoStop}%
\bibitem [{\citenamefont {{McKeever}}\ \emph {et~al.}(2003)\citenamefont
  {{McKeever}}, \citenamefont {{Boca}}, \citenamefont {{Boozer}}, \citenamefont
  {{Buck}},\ and\ \citenamefont {{Kimble}}}]{McKeever03nat}%
  \BibitemOpen
  \bibfield  {author} {\bibinfo {author} {\bibfnamefont {J.}~\bibnamefont
  {{McKeever}}}, \bibinfo {author} {\bibfnamefont {A.}~\bibnamefont {{Boca}}},
  \bibinfo {author} {\bibfnamefont {A.~D.}\ \bibnamefont {{Boozer}}}, \bibinfo
  {author} {\bibfnamefont {J.~R.}\ \bibnamefont {{Buck}}}, \ and\ \bibinfo
  {author} {\bibfnamefont {H.~J.}\ \bibnamefont {{Kimble}}},\ }\href {\doibase
  10.1038/nature01974} {\bibfield  {journal} {\bibinfo  {journal} {Nature}\
  }\textbf {\bibinfo {volume} {425}},\ \bibinfo {pages} {268} (\bibinfo {year}
  {2003})}\BibitemShut {NoStop}%
\bibitem [{\citenamefont {Xie}\ \emph {et~al.}(2007)\citenamefont {Xie},
  \citenamefont {G\"otzinger}, \citenamefont {Fang}, \citenamefont {Cao},\ and\
  \citenamefont {Solomon}}]{Xie07prl}%
  \BibitemOpen
  \bibfield  {author} {\bibinfo {author} {\bibfnamefont {Z.~G.}\ \bibnamefont
  {Xie}}, \bibinfo {author} {\bibfnamefont {S.}~\bibnamefont {G\"otzinger}},
  \bibinfo {author} {\bibfnamefont {W.}~\bibnamefont {Fang}}, \bibinfo {author}
  {\bibfnamefont {H.}~\bibnamefont {Cao}}, \ and\ \bibinfo {author}
  {\bibfnamefont {G.~S.}\ \bibnamefont {Solomon}},\ }\href {\doibase
  10.1103/PhysRevLett.98.117401} {\bibfield  {journal} {\bibinfo  {journal}
  {Phys. Rev. Lett.}\ }\textbf {\bibinfo {volume} {98}},\ \bibinfo {pages}
  {117401} (\bibinfo {year} {2007})}\BibitemShut {NoStop}%
\bibitem [{\citenamefont {{Nomura}}\ \emph {et~al.}(2010)\citenamefont
  {{Nomura}}, \citenamefont {{Kumagai}}, \citenamefont {{Iwamoto}},
  \citenamefont {{Ota}},\ and\ \citenamefont {{Arakawa}}}]{Nomura10natphys}%
  \BibitemOpen
  \bibfield  {author} {\bibinfo {author} {\bibfnamefont {M.}~\bibnamefont
  {{Nomura}}}, \bibinfo {author} {\bibfnamefont {N.}~\bibnamefont {{Kumagai}}},
  \bibinfo {author} {\bibfnamefont {S.}~\bibnamefont {{Iwamoto}}}, \bibinfo
  {author} {\bibfnamefont {Y.}~\bibnamefont {{Ota}}}, \ and\ \bibinfo {author}
  {\bibfnamefont {Y.}~\bibnamefont {{Arakawa}}},\ }\href {\doibase
  10.1038/nphys1518} {\bibfield  {journal} {\bibinfo  {journal} {Nat. Phys.}\
  }\textbf {\bibinfo {volume} {6}},\ \bibinfo {pages} {279} (\bibinfo {year}
  {2010})}\BibitemShut {NoStop}%
\bibitem [{\citenamefont {Schmidt}\ and\ \citenamefont
  {Koch}(2013)}]{Schmidt13}%
  \BibitemOpen
  \bibfield  {author} {\bibinfo {author} {\bibfnamefont {S.}~\bibnamefont
  {Schmidt}}\ and\ \bibinfo {author} {\bibfnamefont {J.}~\bibnamefont {Koch}},\
  }\href {\doibase 10.1002/andp.201200261} {\bibfield  {journal} {\bibinfo
  {journal} {Ann. Phys.}\ }\textbf {\bibinfo {volume} {525}},\ \bibinfo {pages}
  {395} (\bibinfo {year} {2013})}\BibitemShut {NoStop}%
\bibitem [{\citenamefont {Quijandr\'ia}\ \emph {et~al.}(2013)\citenamefont
  {Quijandr\'ia}, \citenamefont {Porras}, \citenamefont {Garc\'ia-Ripoll},\
  and\ \citenamefont {Zueco}}]{Quijandria13}%
  \BibitemOpen
  \bibfield  {author} {\bibinfo {author} {\bibfnamefont {F.}~\bibnamefont
  {Quijandr\'ia}}, \bibinfo {author} {\bibfnamefont {D.}~\bibnamefont
  {Porras}}, \bibinfo {author} {\bibfnamefont {J.~J.}\ \bibnamefont
  {Garc\'ia-Ripoll}}, \ and\ \bibinfo {author} {\bibfnamefont {D.}~\bibnamefont
  {Zueco}},\ }\href {\doibase 10.1103/PhysRevLett.111.073602} {\bibfield
  {journal} {\bibinfo  {journal} {Phys. Rev. Lett.}\ }\textbf {\bibinfo
  {volume} {111}},\ \bibinfo {pages} {073602} (\bibinfo {year}
  {2013})}\BibitemShut {NoStop}%
\bibitem [{Sup()}]{SupMat}%
  \BibitemOpen
  \href@noop {} {}\bibinfo {note} {Supplemental material}\BibitemShut {NoStop}%
\bibitem [{\citenamefont {{Grifoni}}\ and\ \citenamefont
  {{H{\"a}nggi}}(1998)}]{Hanggi_review}%
  \BibitemOpen
  \bibfield  {author} {\bibinfo {author} {\bibfnamefont {M.}~\bibnamefont
  {{Grifoni}}}\ and\ \bibinfo {author} {\bibfnamefont {P.}~\bibnamefont
  {{H{\"a}nggi}}},\ }\href {\doibase 10.1016/S0370-1573(98)00022-2} {\bibfield
  {journal} {\bibinfo  {journal} {Phys. Rep.}\ }\textbf {\bibinfo {volume}
  {304}},\ \bibinfo {pages} {229} (\bibinfo {year} {1998})}\BibitemShut
  {NoStop}%
\bibitem [{\citenamefont {Breuer}\ and\ \citenamefont
  {Petruccione}(2002)}]{Breuer_book}%
  \BibitemOpen
  \bibfield  {author} {\bibinfo {author} {\bibfnamefont {H.-P.}\ \bibnamefont
  {Breuer}}\ and\ \bibinfo {author} {\bibfnamefont {F.}~\bibnamefont
  {Petruccione}},\ }\href@noop {} {\emph {\bibinfo {title} {The theory of open
  quantum systems}}}\ (\bibinfo  {publisher} {Oxford University Press},\
  \bibinfo {year} {2002})\BibitemShut {NoStop}%
\bibitem [{\citenamefont {Mandel}\ and\ \citenamefont
  {Wolf}(1995)}]{Mandel95book}%
  \BibitemOpen
  \bibfield  {author} {\bibinfo {author} {\bibfnamefont {L.}~\bibnamefont
  {Mandel}}\ and\ \bibinfo {author} {\bibfnamefont {E.}~\bibnamefont {Wolf}},\
  }\href@noop {} {\emph {\bibinfo {title} {Optical coherence and quantum
  optics}}}\ (\bibinfo  {publisher} {Cambridge University Press},\ \bibinfo
  {year} {1995})\BibitemShut {NoStop}%
\bibitem [{\citenamefont {Weedbrook}\ \emph {et~al.}(2012)\citenamefont
  {Weedbrook}, \citenamefont {Pirandola}, \citenamefont {Garc\'ia-Patr\'on},
  \citenamefont {Cerf}, \citenamefont {Ralph}, \citenamefont {Shapiro},\ and\
  \citenamefont {Lloyd}}]{Weedbrook12}%
  \BibitemOpen
  \bibfield  {author} {\bibinfo {author} {\bibfnamefont {C.}~\bibnamefont
  {Weedbrook}}, \bibinfo {author} {\bibfnamefont {S.}~\bibnamefont
  {Pirandola}}, \bibinfo {author} {\bibfnamefont {R.}~\bibnamefont
  {Garc\'ia-Patr\'on}}, \bibinfo {author} {\bibfnamefont {N.~J.}\ \bibnamefont
  {Cerf}}, \bibinfo {author} {\bibfnamefont {T.~C.}\ \bibnamefont {Ralph}},
  \bibinfo {author} {\bibfnamefont {J.~H.}\ \bibnamefont {Shapiro}}, \ and\
  \bibinfo {author} {\bibfnamefont {S.}~\bibnamefont {Lloyd}},\ }\href
  {\doibase 10.1103/RevModPhys.84.621} {\bibfield  {journal} {\bibinfo
  {journal} {Rev. Mod. Phys.}\ }\textbf {\bibinfo {volume} {84}},\ \bibinfo
  {pages} {621} (\bibinfo {year} {2012})}\BibitemShut {NoStop}%
\bibitem [{\citenamefont {Scully}\ and\ \citenamefont
  {Zubairy}(1997)}]{Scully97book}%
  \BibitemOpen
  \bibfield  {author} {\bibinfo {author} {\bibfnamefont {M.~O.}\ \bibnamefont
  {Scully}}\ and\ \bibinfo {author} {\bibfnamefont {M.~S.}\ \bibnamefont
  {Zubairy}},\ }\href@noop {} {\emph {\bibinfo {title} {Quantum optics}}}\
  (\bibinfo  {publisher} {Cambridge University Press},\ \bibinfo {year}
  {1997})\BibitemShut {NoStop}%
\bibitem [{\citenamefont {Walls}\ and\ \citenamefont
  {Milburn}(1994)}]{Walls94book}%
  \BibitemOpen
  \bibfield  {author} {\bibinfo {author} {\bibfnamefont {D.~F.}\ \bibnamefont
  {Walls}}\ and\ \bibinfo {author} {\bibfnamefont {G.~J.}\ \bibnamefont
  {Milburn}},\ }\href@noop {} {\emph {\bibinfo {title} {Quantum optics}}}\
  (\bibinfo  {publisher} {Springer},\ \bibinfo {year} {1994})\BibitemShut
  {NoStop}%
\bibitem [{\citenamefont {Cahill}\ and\ \citenamefont
  {Glauber}(1969)}]{Cahill69}%
  \BibitemOpen
  \bibfield  {author} {\bibinfo {author} {\bibfnamefont {K.~E.}\ \bibnamefont
  {Cahill}}\ and\ \bibinfo {author} {\bibfnamefont {R.~J.}\ \bibnamefont
  {Glauber}},\ }\href {\doibase 10.1103/PhysRev.177.1857} {\bibfield  {journal}
  {\bibinfo  {journal} {Phys. Rev.}\ }\textbf {\bibinfo {volume} {177}},\
  \bibinfo {pages} {1857} (\bibinfo {year} {1969})}\BibitemShut {NoStop}%
\bibitem [{\citenamefont {Brune}\ \emph {et~al.}(1992)\citenamefont {Brune},
  \citenamefont {Haroche}, \citenamefont {Raimond}, \citenamefont
  {Davidovich},\ and\ \citenamefont {Zagury}}]{Brune92}%
  \BibitemOpen
  \bibfield  {author} {\bibinfo {author} {\bibfnamefont {M.}~\bibnamefont
  {Brune}}, \bibinfo {author} {\bibfnamefont {S.}~\bibnamefont {Haroche}},
  \bibinfo {author} {\bibfnamefont {J.~M.}\ \bibnamefont {Raimond}}, \bibinfo
  {author} {\bibfnamefont {L.}~\bibnamefont {Davidovich}}, \ and\ \bibinfo
  {author} {\bibfnamefont {N.}~\bibnamefont {Zagury}},\ }\href {\doibase
  10.1103/PhysRevA.45.5193} {\bibfield  {journal} {\bibinfo  {journal} {Phys.
  Rev. A}\ }\textbf {\bibinfo {volume} {45}},\ \bibinfo {pages} {5193}
  (\bibinfo {year} {1992})}\BibitemShut {NoStop}%
\bibitem [{\citenamefont {Garraway}\ and\ \citenamefont
  {Knight}(1992)}]{Garraway92}%
  \BibitemOpen
  \bibfield  {author} {\bibinfo {author} {\bibfnamefont {B.~M.}\ \bibnamefont
  {Garraway}}\ and\ \bibinfo {author} {\bibfnamefont {P.~L.}\ \bibnamefont
  {Knight}},\ }\href {\doibase 10.1103/PhysRevA.46.R5346} {\bibfield  {journal}
  {\bibinfo  {journal} {Phys. Rev. A}\ }\textbf {\bibinfo {volume} {46}},\
  \bibinfo {pages} {R5346} (\bibinfo {year} {1992})}\BibitemShut {NoStop}%
\bibitem [{\citenamefont {Hofheinz}\ \emph {et~al.}(2009)\citenamefont
  {Hofheinz}, \citenamefont {Wang}, \citenamefont {Ansmann}, \citenamefont
  {Bialczak}, \citenamefont {Lucero}, \citenamefont {Neeley}, \citenamefont
  {O'Connell}, \citenamefont {Sank}, \citenamefont {Wenner}, \citenamefont
  {Martinis},\ and\ \citenamefont {Cleland}}]{Hofheinz09nat}%
  \BibitemOpen
  \bibfield  {author} {\bibinfo {author} {\bibfnamefont {M.}~\bibnamefont
  {Hofheinz}}, \bibinfo {author} {\bibfnamefont {H.}~\bibnamefont {Wang}},
  \bibinfo {author} {\bibfnamefont {M.}~\bibnamefont {Ansmann}}, \bibinfo
  {author} {\bibfnamefont {R.~C.}\ \bibnamefont {Bialczak}}, \bibinfo {author}
  {\bibfnamefont {E.}~\bibnamefont {Lucero}}, \bibinfo {author} {\bibfnamefont
  {M.}~\bibnamefont {Neeley}}, \bibinfo {author} {\bibfnamefont {A.~D.}\
  \bibnamefont {O'Connell}}, \bibinfo {author} {\bibfnamefont {D.}~\bibnamefont
  {Sank}}, \bibinfo {author} {\bibfnamefont {J.}~\bibnamefont {Wenner}},
  \bibinfo {author} {\bibfnamefont {J.~M.}\ \bibnamefont {Martinis}}, \ and\
  \bibinfo {author} {\bibfnamefont {A.~N.}\ \bibnamefont {Cleland}},\
  }\href@noop {} {\bibfield  {journal} {\bibinfo  {journal} {Nature}\ }\textbf
  {\bibinfo {volume} {459}},\ \bibinfo {pages} {546} (\bibinfo {year}
  {2009})}\BibitemShut {NoStop}%
\bibitem [{\citenamefont {Koch}\ \emph {et~al.}(2007)\citenamefont {Koch},
  \citenamefont {Yu}, \citenamefont {Gambetta}, \citenamefont {Houck},
  \citenamefont {Schuster}, \citenamefont {Majer}, \citenamefont {Blais},
  \citenamefont {Devoret}, \citenamefont {Girvin},\ and\ \citenamefont
  {Schoelkopf}}]{Koch07}%
  \BibitemOpen
  \bibfield  {author} {\bibinfo {author} {\bibfnamefont {J.}~\bibnamefont
  {Koch}}, \bibinfo {author} {\bibfnamefont {T.~M.}\ \bibnamefont {Yu}},
  \bibinfo {author} {\bibfnamefont {J.}~\bibnamefont {Gambetta}}, \bibinfo
  {author} {\bibfnamefont {A.~A.}\ \bibnamefont {Houck}}, \bibinfo {author}
  {\bibfnamefont {D.~I.}\ \bibnamefont {Schuster}}, \bibinfo {author}
  {\bibfnamefont {J.}~\bibnamefont {Majer}}, \bibinfo {author} {\bibfnamefont
  {A.}~\bibnamefont {Blais}}, \bibinfo {author} {\bibfnamefont {M.~H.}\
  \bibnamefont {Devoret}}, \bibinfo {author} {\bibfnamefont {S.~M.}\
  \bibnamefont {Girvin}}, \ and\ \bibinfo {author} {\bibfnamefont {R.~J.}\
  \bibnamefont {Schoelkopf}},\ }\href {\doibase 10.1103/PhysRevA.76.042319}
  {\bibfield  {journal} {\bibinfo  {journal} {Phys. Rev. A}\ }\textbf {\bibinfo
  {volume} {76}},\ \bibinfo {pages} {042319} (\bibinfo {year}
  {2007})}\BibitemShut {NoStop}%
\bibitem [{\citenamefont {Schreier}\ \emph {et~al.}(2008)\citenamefont
  {Schreier}, \citenamefont {Houck}, \citenamefont {Koch}, \citenamefont
  {Schuster}, \citenamefont {Johnson}, \citenamefont {Chow}, \citenamefont
  {Gambetta}, \citenamefont {Majer}, \citenamefont {Frunzio}, \citenamefont
  {Devoret}, \citenamefont {Girvin},\ and\ \citenamefont
  {Schoelkopf}}]{Schreier08}%
  \BibitemOpen
  \bibfield  {author} {\bibinfo {author} {\bibfnamefont {J.~A.}\ \bibnamefont
  {Schreier}}, \bibinfo {author} {\bibfnamefont {A.~A.}\ \bibnamefont {Houck}},
  \bibinfo {author} {\bibfnamefont {J.}~\bibnamefont {Koch}}, \bibinfo {author}
  {\bibfnamefont {D.~I.}\ \bibnamefont {Schuster}}, \bibinfo {author}
  {\bibfnamefont {B.~R.}\ \bibnamefont {Johnson}}, \bibinfo {author}
  {\bibfnamefont {J.~M.}\ \bibnamefont {Chow}}, \bibinfo {author}
  {\bibfnamefont {J.~M.}\ \bibnamefont {Gambetta}}, \bibinfo {author}
  {\bibfnamefont {J.}~\bibnamefont {Majer}}, \bibinfo {author} {\bibfnamefont
  {L.}~\bibnamefont {Frunzio}}, \bibinfo {author} {\bibfnamefont {M.~H.}\
  \bibnamefont {Devoret}}, \bibinfo {author} {\bibfnamefont {S.~M.}\
  \bibnamefont {Girvin}}, \ and\ \bibinfo {author} {\bibfnamefont {R.~J.}\
  \bibnamefont {Schoelkopf}},\ }\href {\doibase 10.1103/PhysRevB.77.180502}
  {\bibfield  {journal} {\bibinfo  {journal} {Phys. Rev. B}\ }\textbf {\bibinfo
  {volume} {77}},\ \bibinfo {pages} {180502} (\bibinfo {year}
  {2008})}\BibitemShut {NoStop}%
\bibitem [{\citenamefont {Diehl}\ \emph {et~al.}(2008)\citenamefont {Diehl},
  \citenamefont {Micheli}, \citenamefont {Kantian}, \citenamefont {Kraus},
  \citenamefont {Buchler},\ and\ \citenamefont {Zoller}}]{Diehl08}%
  \BibitemOpen
  \bibfield  {author} {\bibinfo {author} {\bibfnamefont {S.}~\bibnamefont
  {Diehl}}, \bibinfo {author} {\bibfnamefont {A.}~\bibnamefont {Micheli}},
  \bibinfo {author} {\bibfnamefont {A.}~\bibnamefont {Kantian}}, \bibinfo
  {author} {\bibfnamefont {B.}~\bibnamefont {Kraus}}, \bibinfo {author}
  {\bibfnamefont {H.~P.}\ \bibnamefont {Buchler}}, \ and\ \bibinfo {author}
  {\bibfnamefont {P.}~\bibnamefont {Zoller}},\ }\href {\doibase
  10.1038/nphys1073} {\bibfield  {journal} {\bibinfo  {journal} {Nat. Phys.}\
  }\textbf {\bibinfo {volume} {4}},\ \bibinfo {pages} {878} (\bibinfo {year}
  {2008})}\BibitemShut {NoStop}%
\bibitem [{\citenamefont {Verstraete}\ \emph {et~al.}(2009)\citenamefont
  {Verstraete}, \citenamefont {Wolf},\ and\ \citenamefont
  {Cirac}}]{Verstraete09}%
  \BibitemOpen
  \bibfield  {author} {\bibinfo {author} {\bibfnamefont {F.}~\bibnamefont
  {Verstraete}}, \bibinfo {author} {\bibfnamefont {M.~M.}\ \bibnamefont
  {Wolf}}, \ and\ \bibinfo {author} {\bibfnamefont {J.~I.}\ \bibnamefont
  {Cirac}},\ }\href {\doibase 10.1038/nphys1342} {\bibfield  {journal}
  {\bibinfo  {journal} {Nat. Phys.}\ }\textbf {\bibinfo {volume} {5}},\
  \bibinfo {pages} {633} (\bibinfo {year} {2009})}\BibitemShut {NoStop}%
\bibitem [{\citenamefont {Barreiro}\ \emph {et~al.}(2010)\citenamefont
  {Barreiro}, \citenamefont {Schindler}, \citenamefont {Guhne}, \citenamefont
  {Monz}, \citenamefont {Chwalla}, \citenamefont {Roos}, \citenamefont
  {Hennrich},\ and\ \citenamefont {Blatt}}]{Barreiro10}%
  \BibitemOpen
  \bibfield  {author} {\bibinfo {author} {\bibfnamefont {J.~T.}\ \bibnamefont
  {Barreiro}}, \bibinfo {author} {\bibfnamefont {P.}~\bibnamefont {Schindler}},
  \bibinfo {author} {\bibfnamefont {O.}~\bibnamefont {Guhne}}, \bibinfo
  {author} {\bibfnamefont {T.}~\bibnamefont {Monz}}, \bibinfo {author}
  {\bibfnamefont {M.}~\bibnamefont {Chwalla}}, \bibinfo {author} {\bibfnamefont
  {C.~F.}\ \bibnamefont {Roos}}, \bibinfo {author} {\bibfnamefont
  {M.}~\bibnamefont {Hennrich}}, \ and\ \bibinfo {author} {\bibfnamefont
  {R.}~\bibnamefont {Blatt}},\ }\href {\doibase 10.1038/nphys1781} {\bibfield
  {journal} {\bibinfo  {journal} {Nat. Phys.}\ }\textbf {\bibinfo {volume}
  {6}},\ \bibinfo {pages} {943} (\bibinfo {year} {2010})}\BibitemShut {NoStop}%
\bibitem [{\citenamefont {Barreiro}\ \emph {et~al.}(2011)\citenamefont
  {Barreiro}, \citenamefont {Muller}, \citenamefont {Schindler}, \citenamefont
  {Nigg}, \citenamefont {Monz}, \citenamefont {Chwalla}, \citenamefont
  {Hennrich}, \citenamefont {Roos}, \citenamefont {Zoller},\ and\ \citenamefont
  {Blatt}}]{Barreiro11}%
  \BibitemOpen
  \bibfield  {author} {\bibinfo {author} {\bibfnamefont {J.~T.}\ \bibnamefont
  {Barreiro}}, \bibinfo {author} {\bibfnamefont {M.}~\bibnamefont {Muller}},
  \bibinfo {author} {\bibfnamefont {P.}~\bibnamefont {Schindler}}, \bibinfo
  {author} {\bibfnamefont {D.}~\bibnamefont {Nigg}}, \bibinfo {author}
  {\bibfnamefont {T.}~\bibnamefont {Monz}}, \bibinfo {author} {\bibfnamefont
  {M.}~\bibnamefont {Chwalla}}, \bibinfo {author} {\bibfnamefont
  {M.}~\bibnamefont {Hennrich}}, \bibinfo {author} {\bibfnamefont {C.~F.}\
  \bibnamefont {Roos}}, \bibinfo {author} {\bibfnamefont {P.}~\bibnamefont
  {Zoller}}, \ and\ \bibinfo {author} {\bibfnamefont {R.}~\bibnamefont
  {Blatt}},\ }\href {\doibase 10.1038/nature09801} {\bibfield  {journal}
  {\bibinfo  {journal} {Nature}\ }\textbf {\bibinfo {volume} {470}},\ \bibinfo
  {pages} {486} (\bibinfo {year} {2011})}\BibitemShut {NoStop}%
\bibitem [{\citenamefont {Krauter}\ \emph {et~al.}(2011)\citenamefont
  {Krauter}, \citenamefont {Muschik}, \citenamefont {Jensen}, \citenamefont
  {Wasilewski}, \citenamefont {Petersen}, \citenamefont {Cirac},\ and\
  \citenamefont {Polzik}}]{Krauter11}%
  \BibitemOpen
  \bibfield  {author} {\bibinfo {author} {\bibfnamefont {H.}~\bibnamefont
  {Krauter}}, \bibinfo {author} {\bibfnamefont {C.~A.}\ \bibnamefont
  {Muschik}}, \bibinfo {author} {\bibfnamefont {K.}~\bibnamefont {Jensen}},
  \bibinfo {author} {\bibfnamefont {W.}~\bibnamefont {Wasilewski}}, \bibinfo
  {author} {\bibfnamefont {J.~M.}\ \bibnamefont {Petersen}}, \bibinfo {author}
  {\bibfnamefont {J.~I.}\ \bibnamefont {Cirac}}, \ and\ \bibinfo {author}
  {\bibfnamefont {E.~S.}\ \bibnamefont {Polzik}},\ }\href {\doibase
  10.1103/PhysRevLett.107.080503} {\bibfield  {journal} {\bibinfo  {journal}
  {Phys. Rev. Lett.}\ }\textbf {\bibinfo {volume} {107}},\ \bibinfo {pages}
  {080503} (\bibinfo {year} {2011})}\BibitemShut {NoStop}%
\end{thebibliography}%

\end{document}